\documentclass[journal=jacsat,manuscript=article]{achemso}

\usepackage[version=3]{mhchem} % Formula subscripts using \ce{}
\usepackage{afterpage} 
\usepackage{amsmath}
\usepackage{csquotes}

\author{Rana Bachnak}
\affiliation[UMN ME]
{Department of Mechanical Engineering, University of Minnesota - Twin Cities, Minneapolis, MN}
\author{Cari S. Dutcher}
\affiliation[UMN ME]
{Department of Mechanical Engineering, University of Minnesota - Twin Cities, Minneapolis, MN}
\alsoaffiliation[UMN CHEM]{Department of Chemical Engineering and Material Science, University of Minnesota - Twin Cities, Minneapolis, MN}
\email{cdutcher@umn.edu}
\title[An \textsf{achemso} demo]
  {Unlocking Surfactant Synergies for Fluorine Replacements: A Microfluidic Exploration of Emulsion Stability and Coalescence Dynamics}

\abbreviations{IR,NMR,UV}
\begin{document}

%%%%%%%%%%%%%%%%%%%%%%%%%%%%%%%%%%%%%%%%%%%%%%%%%%%%%%%%%%%%%%%%%%%%%

\begin{abstract}
Fluorinated surfactants are widely utilized in various applications, including firefighting foams, coatings, and lubricants. Their wide utility is due to their unique properties, such as their ability to effectively lower surface/interfacial tension. However, the excessive use of fluorinated surfactants has raised environmental and health concerns due to their persistence in the environment and potential toxicity. The need for effective and sustainable surfactant replacements has therefore become critical. Our study investigates alternative surfactants mixtures that have been proposed to replicate the performance of traditional fluorinated surfactants, while minimizing environmental impact. Here, we study the stability of thin aqueous liquid bridges found in water in oil emulsions, with surfactant mixtures shown to provide synergistic fire suppression in fluorine-free fire fighting foams. Binary surfactant-water solutions of TritonX-100, Glucopon 215, Dow 502W were used, in addition to a ternary surfactant-surfactant-water mixtures of Glucopon 215 and Dow502W. Emulsion stability is assessed using a microfluidic device designed to measure the coalescence frequency of an emulsion. Coalescence frequency is measured as a function of total flow rate and surfactant concentration. A non-monotonic trend of coalescence frequency with flow rate is found for all the surfactants used. In addition, a non-dimensional frequency is introduced and is found to decrease with the Capillary number. The coalescence frequency of the surfactant mixture is found to be close to that of Dow502W, suggesting a synergistic behavior between Glucopon215 and Dow502W, even with less of the siloxane surfactant present in the formulation. This behavior highlights the potential for pairing slower hydrocarbon surfactants with faster siloxane surfactants to achieve an effective alternative stabilizing agent for fluorine free formulation. 
\end{abstract}

\section*{Keywords}
Coalescence, frequency, emulsions, surfactant mixtures, fluorine replacements

\section{Introduction}

Emulsions and foams play a crucial role in several chemical and engineering industries, making their stability a critical factor for optimal performance and functionality \cite{prud2017foams}. Foams and emulsions are mainly stabilized by surfactants, which reduce their surface/interfacial tension \cite{briceno2021aqueous,langevin2023recent}. Different types of surfactants can be used, depending on the application, including hydrocarbon, fluorinated, and silicon-based surfactants \cite{matusiak2020influence,zhao2024research}. There are various similarities in the mechanisms underlying the surfactant-enabled stability of foams and emulsions, which makes the study of one insightful and applicable to the understanding of both \cite{hunter2008role}. 

Fluorinated surfactants, with partially or completely fluorinated hydrophobic tail, possess unique properties that make them suitable in different applications, such as in fire-fighting foams, paints, coatings, and drug delivery \cite{krafft2001fluorocarbons,czajka2015surfactants}. The high electro-negativity of fluorine increases the chemical stability of fluorinated surfactants, compared to hydrocarbon ones \cite{hussain2022fluorinated}. Fluorinated surfactants are more effective in reducing surface/interfacial tension than hydrocarbon surfactants with similar tail lengths. These unique properties make fluorinated surfactants useful in a wide range of applications involving surfactant-stabilized emulsions, such as in lubricants, cosmetics, coatings, paints, and adhesives \cite{pabon2002fluorinated,porter2013handbook,czajka2015surfactants,hussain2022fluorinated}.

However, concerns about the environmental persistence and potential health impacts of per- and polyfluoroalkyl (PFAS) have led to increased research into fluorine-free surfactant alternatives \cite{bell2021exposure,sunderland2019review,pelch2019pfas,hetzer2014fire,kaller2023evaluation} . To explore the use of fluorine-free alternatives for emulsion stability, we will examine a foam-based application that traditionally relies on fluorinated surfactants: firefighting foams, to inform a possible fluorinated surfactant replacement. 

Fuel fires pose a significant threat to safety, particularly in industrial settings, transportation hubs, and energy facilities. These fires can rapidly spread and intensify, due to the high energy density of liquid fuels, and are challenging to control and extinguish \cite{islam2024understanding,benali2021understanding}. The complexity and danger of these fires necessitate specialized firefighting techniques and equipment. Aqueous film forming foams (AFFF), containing fluorinated surfactants, have been commonly used for fire suppression \cite{hinnant2018analytically,ananth2019synergisms,moody2000perfluorinated,buck2011perfluoroalkyl} . Those film-forming systems are designed to create a barrier between the fuel and oxygen, effectively smothering the fire \cite{islam2024understanding}.

The development of effective fluorine-free firefighting foams involves careful consideration of surfactant properties, including foamability, foam stability, and fire suppression performance. An effective surfactant formulation would be one that can match the performance of fluorinated foams while minimizing environmental impact. The combination of surfactants from different chemical classes has shown promise in enhancing performance and can hold significant potential as key alternatives to replace PFAS-based surfactants \cite{bera2013synergistic,del2007mixed,kume2008review}, when compared to a reference aqueous film forming foam (AFFF) formulation \cite{hinnant2018analytically}. Simply replacing fluorinated surfactants with a fluorine-free surfactant in AFFF formulations was not sufficient, but mixtures of replacement surfactants showed promise. For example, Ananth \textit{et al.} \cite{ananth2019synergisms} reported the effectiveness of a mixture of a siloxane-polyoxyethylene and an alkyl polyglycoside surfactant in suppressing a heptane pool fire, by measuring the foam degradation, and fuel transport rates. The fire-extinction performance was tested on
both small and large scales, showing consistent results. The surfactant mixtures achieved comparable fire suppression performance to traditional fluorosurfactant-based foams, extinguishing heptane pool fires in a manner approaching that of commercial AFFF formulations. The study demonstrates a remarkable synergistic effect when combining Glucopon 215 (G215) and Dow502W siloxane surfactant in fire suppression. Individually, neither G215 nor Dow502W could extinguish the fire at reasonable foam flow rates, requiring exceedingly high rates of 2100 \(\text{mL/min}\) (74 \(\text{L/m}^2/\text{min}\)) and 1550 \(\text{mL/min}\) (54.7 \(\text{L/m}^2/\text{min}\)), respectively, to even approach effectiveness. However, when the two surfactants were combined in a 3:2 ratio, the foam extinguished the fire at a significantly reduced foam flow rate of 453 \(\text{mL/min}\) (16 \(\text{L/m}^2/\text{min}\)) within 197 seconds. This composition highlights a strong synergistic interaction, greatly enhancing fire suppression performance compared to the individual components. The study showed that by optimizing the surfactant composition, siloxane and glycoside surfactant mixtures could serve as potential fluorine-free alternatives to conventional AFFF. 

In addition, Ananth \textit{et al.} \cite{ananth2019synergisms} highlighted the synergistic effects between G215 and Dow502W in their enhanced foam stability compared to the individual surfactants. The surfactant mixture exhibited a smaller foam degradation rate than either surfactant alone. For instance, foams generated using a formulation containing either of G215 and Dow502W surfactants degraded completely in 240 and 480 seconds, respectively. However, when combined in a formulation containing G215 and Dow502W, the foam exhibited a significantly longer degradation time than each of the individual surfactants (900 seconds). While these results clearly demonstrate synergistic effects in foam stability, it is uncertain whether similar synergism will be observed in other applications beyond foams, such as emulsion systems, necessitating further investigation.

Microfluidic approaches provide a powerful platform for investigating the stability of emulsions and foams due to their ability to precisely control and manipulate small droplets or bubbles, at length scales relevant to emulsions and foams, in well-defined environments. These systems allow for the formation of uniform droplets or bubbles and monitor their behavior under controlled conditions, such as varying flow rates, surfactant types, and surfactant concentrations\cite{krebs2012microfluidic,baret2009kinetic,bremond2008decompressing,krebs2013coalescence,tan2004design}. In microfluidics, droplet stability is often measured as the droplet resistance to coalescence \cite{narayan2020insights,narayan2022correction,bachnak2024influence}. Coalescence can be studied either for a pair of droplets \cite{bremond2008decompressing,bachnak2024influence,bachnak2023effect}, or for a droplet ensemble \cite{dudek2020microfluidic,baret2009kinetic}. When studying coalescence of a pair of droplets, stability is quantified by measuring the film drainage time, where a higher time indicates more stable droplets \cite{bachnak2023effect}. As for droplet ensembles, stability is measured by finding coalescence frequencies, where a lower frequency indicates a more stable emulsion \cite{dudek2020microfluidic}.

In this study, we investigate emulsion stability by examining the coalescence behavior of an entire droplet ensemble, rather than isolated droplet pairs, to provide a more comprehensive understanding of emulsion stability. We study the innovative synergistic interactions between Dow502W and Glucopon 215 in stabilizing emulsions. The effectiveness is measured by studying the stability of a water-in-oil emulsion containing those surfactants. The stability is measured as the resistance of the thin liquid bridges to coalescence. The interfacial properties of the surfactants are first measured using the pendant drop method, followed by evaluating the emulsion stability using a microfluidic device designed to measure coalescence frequency of an emulsion. The method for measuring coalescence frequency is first tested with TritonX-100, as it is well-documented in the literature \cite{chen_under_revision} , exhibits a rapid rate of IFT decay that simplifies its application, and is water-soluble, making it an ideal candidate for preliminary validation. The coalescence frequency obtained for surfactant mixtures, is compared to that obtained in the case of individual surfactants, to provide insight into how mixtures that were effective in stabilizing fire-fighting foams contribute to emulsion stabilization. This study provides valuable insights into emulsion stability, which is crucial in applications that rely on fluorinated surfactants for effective stabilization, such as lubricant and coating applications \cite{basset1984oil,hu2017tribological}. By investigating emulsion stability in the presence of surfactant mixtures, the study provides valuable guidance on potential fluorine-free surfactant replacements.

\section{Materials and Methods}

\subsection{Chemicals}

Interfacial tension and coalescence measurements are performed for water in oil systems in the presence of surfactants. Light mineral oil (LMO) (Sigma-Aldrich) is
used as the continuous phase, having a viscosity of 26.8 mPa.s and a density of 0.827 g. LMO is selected for this study due to its extensive characterization in the literature, its ease of use as a model oil, and its well-documented properties. HPLC grade water (Fisher) is used as
the dispersed phase. Three types of water-soluble surfactants are used: TritonX-100, Glucopon 215 CS UP, which will be referred to as G215 (an alkyl polyglycoside concentrate contributed by BASF Corporation, Ludwigshafen, Germany) \cite{hinnant2018analytically}, and Dow Corning 502W Additive, which will be referred to as Dow502W (a silicone polyether copolymer, a 100\% by weight concentrate contributed by Dow \cite{ananth2023development}. In addition, a mixture of 3:2 by weight of Dow502W:G215 was also used, while keeping the same total surfactant concentration as that used with the individual surfactant cases. A wide range of TritonX-100 concentrations were used: 100, 150, 200, 250, 300, and 400 ppm. For G215, Dow502W, and the mixture, different  concentrations were also used: 40, 80, 120, 160, and 200 ppm. All the concentrations used were below the critical micelle concentration (CMC), except for TritonX-100 at 400ppm, and Dow502W at 200ppm, which is its CMC. 

\subsection{Interfacial Tension Measurements}
The pendant drop method is used to determine the interfacial tension of water-in-oil in the presence of water-soluble suractants: G215, Dow502W, and the surfactant mixture. The interfacial tension measurements in the presence of TritonX-100 are reported in a study currently under revision \cite{chen_under_revision} . In the pendant drop method, a water droplet containing the surfactants is suspended at the tip of a needle within the surrounding LMO. The shape of the droplet, which results from the balance between gravity and surface forces, is analyzed to determine the interfacial tension. The drop assumes a characteristic shape, influenced by the interfacial tension and the density difference between the two fluids. Using precise image analysis and fitting the droplet profile to the Young-Laplace equation, the interfacial tension can be accurately quantified \cite{berry2015measurement,narayan2018removing} . In addition to measuring the IFT using the pendant drop method, the IFT is also measured using a microfluidic method previously reported in literature \cite{narayan2018removing, bachnak2024influence}. The microfluidic IFT measurement is performed for the surfactant exhibiting the slowest IFT decay rate, as will be found in the \enquote{Interfacial Tension Results} section to ensure that the IFT has reached equilibrium within the microfluidic coalescence device, which is detailed in the following section.

\subsection{Microfluidic Device Design}
Microfluidics offers significant advantages for studying droplet coalescence due to their precise control over flow conditions, droplet size, and droplet spacing. The device used in this study draws on the body of work that studies emulsions dynamics and droplet coalescence in microfluidic devices, both for isolated binary droplets \cite{narayan2020insights,bachnak2024influence,bremond2008decompressing,bachnak2023effect} and droplet ensembles \cite{krebs2012microfluidic,dudek2020microfluidic,baret2009kinetic}.  In this work, we build on that literature by using a similar device for measuring coalescence frequency, which includes an expansion chamber to slow down the droplets and increase their coalescence probability. However, we introduce different methods for analyzing the coalescence frequency, as will be discussed in following sections: \enquote{Coalescence Frequency Calculation}, \enquote{Coalescence Frequency Results}, and \enquote{Non-Dimensional Analysis of Coalescence Frequency}.

\begin{figure}[H]
   \includegraphics[width=\textwidth]{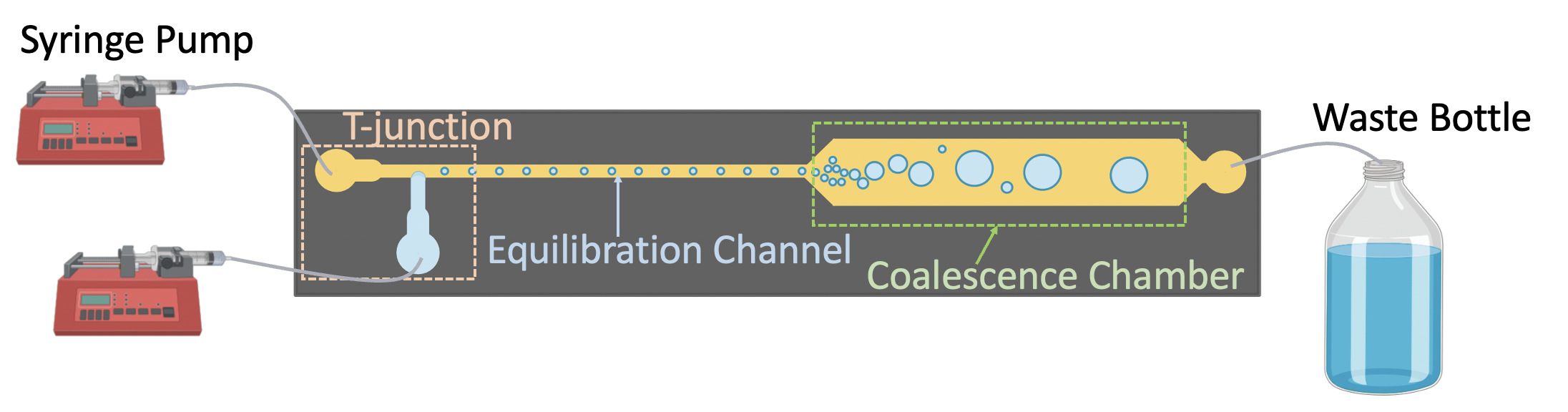} 
  \caption{Schematic of the microfluidic device design used to measure coalescence frequency. Syringe pumps are used to drive the water and oil phases into the device. The figure was created in BioRender. Bachnak, R. (2025) https://BioRender.com/d37m715}
  \label{MicrofluidicDevice}
\end{figure}

The microfluidic device used in this study to measure coalescence frequency, as illustrated in Figure \ref{MicrofluidicDevice} consists of a T-junction used to form droplets on-chip. The droplets then travel through an \enquote{Equilibration Channel}, used to make enough time for the surfactants to absorb to the droplet interface, and allow the IFT to reach a value close to equilibrium, before entering the coalescence chamber. At the entry region of the coalescence chamber, the droplets collect together to form an emulsion, and they start coalescing as they flow through the coalescence chamber. The microfluidic device was fabricated using a 150-micrometer thick SUEX sheet through photolithography in a cleanroom. The SUEX was laminated onto a silicon wafer, soft baked, and then patterned using UV exposure through a photomask.

The experimental setup consists of the microfluidic device mounted on an inverted microscope (IX83), and a high speed camera (Photron Mini UX100) used to take video recordings of the emulsions in the device at a frame rate of 500 fps. Syringe pumps were employed to precisely drive the continuous and dispersed phases into the microfluidic device at a T-junction, where water served as the dispersed phase and oil as the continuous phase. Six different total flow rates were tested, maintaining a constant ratio between the flow rates of the dispersed and continuous phases. This consistent ratio ensured that the droplet size remained uniform across all experiments, allowing for accurate comparison of coalescence behavior under varying flow conditions, as well as varying surfactant types and concentrations.

\subsection{Coalescence Frequency Calculation}
\begin{figure}[H]
   \includegraphics[width=\textwidth]{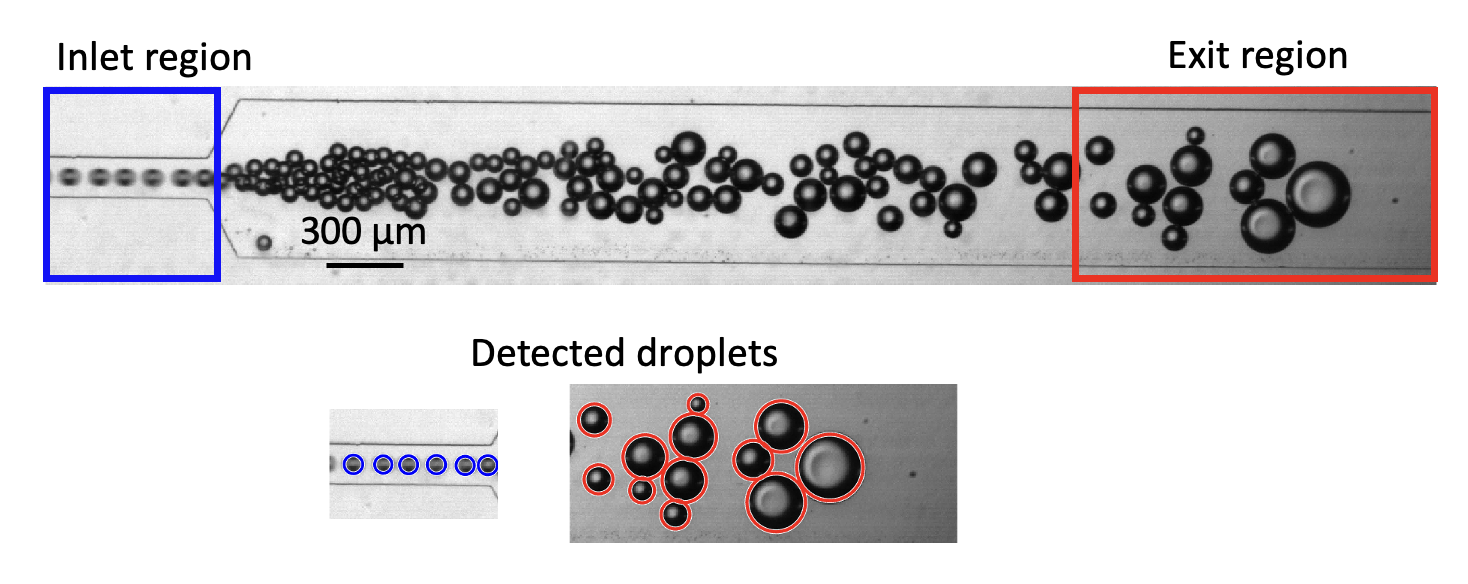} 
  \caption{Example of droplets detected at the inlet and outlet region of the coalescence chamber.}
  \label{ImageAnalysis}
\end{figure}

In order to quantify emulsion stability, the coalescence frequency is used. After recording videos of the coalescence events, the videos are analyzed using MATLAB. Droplet volumes are analyzed at inlet and outlet regions of the device, using the droplet size. The droplet diameters are first found and compared with the height of the channel. For diameter less than 150 $\mu$m, the volume is calculated using the volume of a sphere. For diameters greater than 150 $\mu$m, which typically occurs at the outlet of the channel, the volume is calculated using Pappus's theorem \cite{gunjan2021cloaked} .The inlet region is picked just before the droplets enter the coalescence chamber.  An example of droplets coalescing in the microfluidic device, with droplets detected in the inlet and outlet regions is illustrated in Figure \ref{ImageAnalysis}. The total number of coalescence events is identified by comparing the number and volume of droplets at the inlet and outlet. Specifically, a coalescence event is recognized when two or more smaller droplets at the inlet merge into a single larger droplet at the outlet. In previous studies \cite{krebs2012microfluidic,dudek2020microfluidic,baret2009kinetic}, coalescence frequency was typically calculated based on the number of coalescence events involving a single droplet, by finding the average volume of droplets at the exit region. In our approach, we used the total number of coalescence events to provide a more comprehensive representation of system-wide interactions and to capture collective behavior under varying experimental conditions. The number of coalescence events is calculated as:

\begin{equation}
N_C = \frac{V_{\text{total\_exit}}}{V_{\text{droplet\_inlet}}} - N_{exit} = (V_{ratio} - 1) * N_{exit}
\label{1}
\end{equation} 

Where $V_{\text{total\_exit}}$ is the total volume of droplets in the exit region, $V_{\text{droplet\_inlet}}$ is the average volume of a single droplet in the inlet region, $N_{exit}$ is the number of droplets in the exit region, and $V_{ratio}$ is the average volume ratio of a droplet between the exit and the inlet regions.

After finding the number of coalescence events, the coalescence frequency is found as:

\begin{equation}
f = \frac{N_C}{t_{\text{res}}} 
\label{2}
\end{equation} 

Where $t_{\text{res}}$ is the residence time of the droplet, or the time it takes to travel from the inlet to the exit region, whose locations are fixed.

\section{Results and discussion}

\subsection{Interfacial Tension Results}

\begin{figure}[H]
   \includegraphics[width=\textwidth]{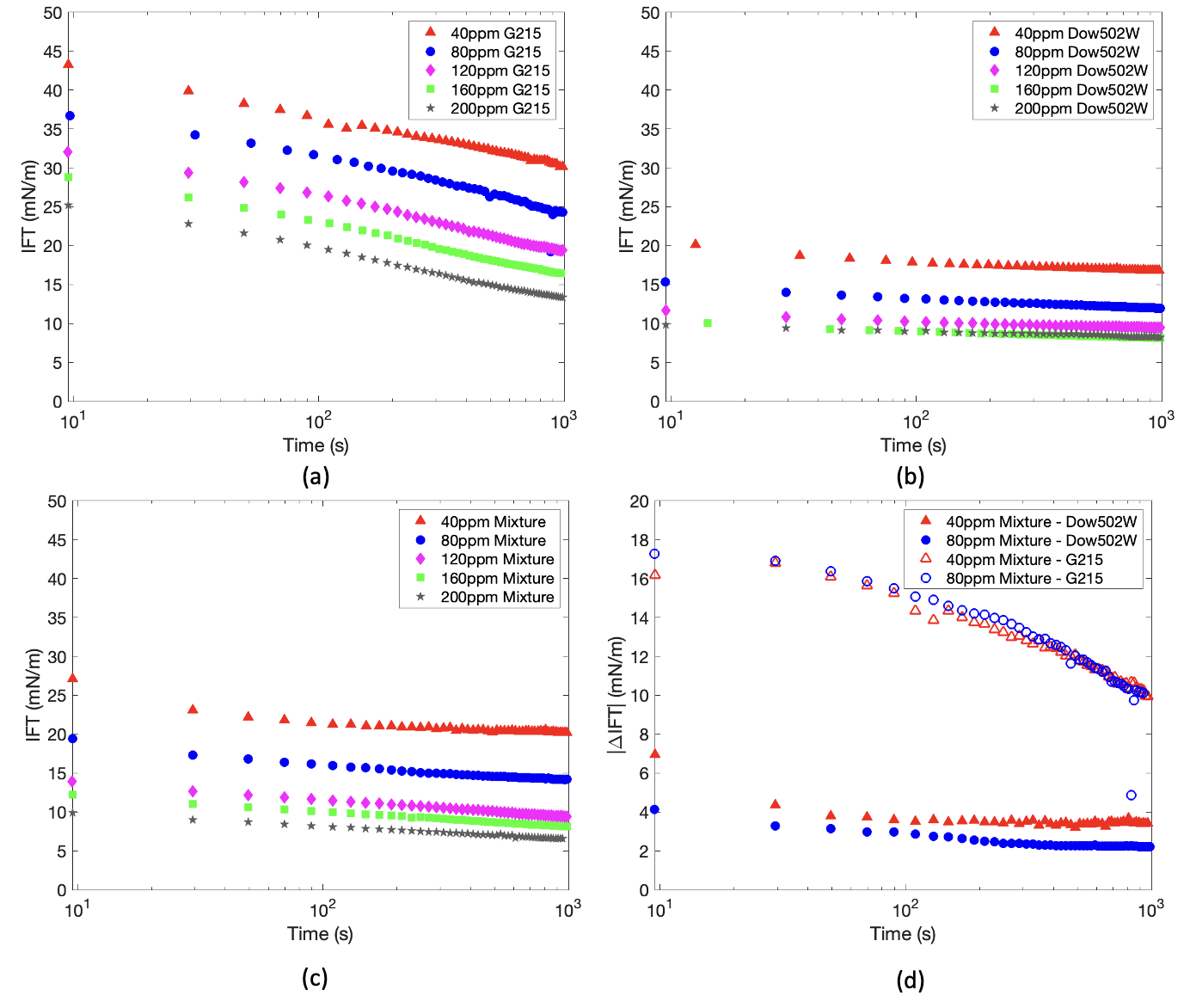} 
  \caption{IFT as a function of time for water in LMO droplets in the presence of (a) G215, (b) Dow502W, and (c) 3:2Dow502:G215 wt ratio mixture, as well as (d) difference between the IFT values of the surfactant mixture and each of G215 and Dow502W at 40 and 80ppm concnetrations. Note that the plots presented here are based on data averaged over 20 consecutive points. The actual values of IFT as a function of time are presented in Figure SI1.}
  \label{Pendant}
\end{figure}

In the pendant drop experiments, interfacial tension measurements were performed using two types of surfactants, G215 and Dow502W, as well as their mixture, across various concentrations. The results, plotted in Figure \ref{Pendant} show a clear trend of decreasing IFT with increasing surfactant concentration for all cases, consistent with the typical behavior of surfactants adsorbing at the interface and reducing interfacial tension. G215 exhibited a relatively slow decay in IFT with time, as shown in Figure \ref{Pendant}(a) compared to Dow502W (Figure \ref{Pendant}(b)), indicating a more gradual adsorption or lower surface activity. In contrast,  Dow502W demonstrated a much steeper reduction in IFT, suggesting faster adsorption kinetics or higher surface activity. Interestingly, the surfactant mixture followed the trend of the faster surfactant Dow502W, with the IFT decreasing rapidly with concentration (Figure \ref{Pendant}(c)), likely due to the dominance of Dow502W in controlling the interfacial behavior in the mixture. The difference in IFT values between the mixture and each of the two individual surfactants is more clearly illustrated in Figure \ref{Pendant}(d). The plot shows the absolute difference in IFT values between the mixture and each of G215 and Dow502W at 40 ppm. Notably, the difference is significantly smaller between the mixture and Dow502W compared to the difference between the mixture and G215 These results highlight the importance of surfactant type and concentration in determining interfacial properties and suggest that the faster-acting surfactant can play a critical role in mixture behavior.

\subsection{Coalescence Frequency Results}
The first step in performing coalescence measurements is to make sure that the IFT reaches equilibrium before the droplets enter the coalescence chamber. To achieve this, microfluidic IFT measurements are done for water droplets with G215 at its lowest concentration (40ppm) in LMO, using the microfluidic tensiometer method outlined in the study of Narayan \textit{et al.} \cite{narayan2018removing}. The G215 system at a 40 ppm concentration is used here as it has the slowest decay rate in IFT among all the surfactants used and at all concentrations. This implies that if this system is at equilibrium, all the others are at equilibrium. The plot of IFT as a function of time for microfluidics is included in Figure SI2. The IFT value reaches equilibrium at the level of the first tensiometer, within at most 1.37 s. The flow rate of the coalescence frequency varies between 13 and 91 $\mu$L/min. At the highest flow rate used, the droplets enter the coalescence chamber 0.6 s after their generation, which is estimated to provide sufficient time for the IFT to reach equilibrium.

\begin{figure}[H]
   \includegraphics[width=\textwidth]{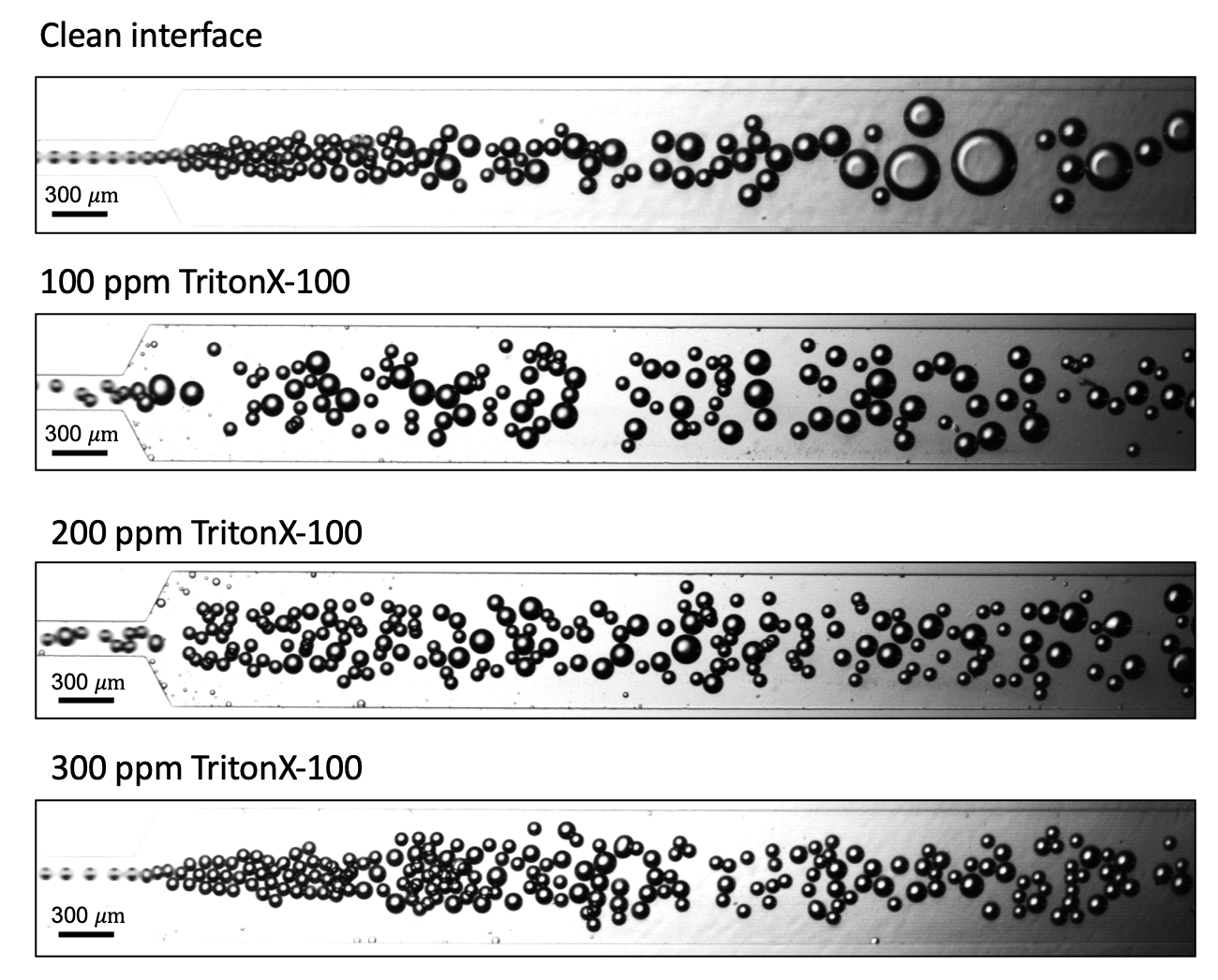} 
  \caption{Visualization of coalescence events in the coalescence chamber, illustrating the progression from a clean interface to varying TritonX-100 concentrations, highlighting the influence of surfactants on coalescence dynamics.}
  \label{Triton Mosaic}
\end{figure}

To visualize the effect of surfactant concentration on the coalescence frequency, images of the coalescence chamber are taken, for the case of TritonX-100, at different surfactant, as well as for the clean interface, as shown in Figure \ref{Triton Mosaic}. It appears the stability of the emulsion improves with the addition of Triton X-100, and this effect becomes more pronounced as its concentration increases. This initial finding is based on a visual observation of the droplets in the outlet region, where their volume visibly decreases with increasing surfactant concentration. This qualitative observation suggests reduced coalescence and improved stability. However, frequency calculations must be done to further confirm this observation.

The analysis of droplet coalescence frequency is complex, even when the same flow rate ratio of dispersed to continuous phase is maintained. This complexity arises because the rate of droplet formation and the droplet sizes can vary due to changes interfacial dynamics, even at a fixed total flow rate. Additionally, the droplets can break up during flow, introducing further variability. The residence time of droplets in the channel also varies depending on the total flow rate. These factors collectively influence the frequency of coalescence, requiring careful consideration and incorporation into the analysis.

\begin{figure}[H]
   \includegraphics[width=\textwidth]{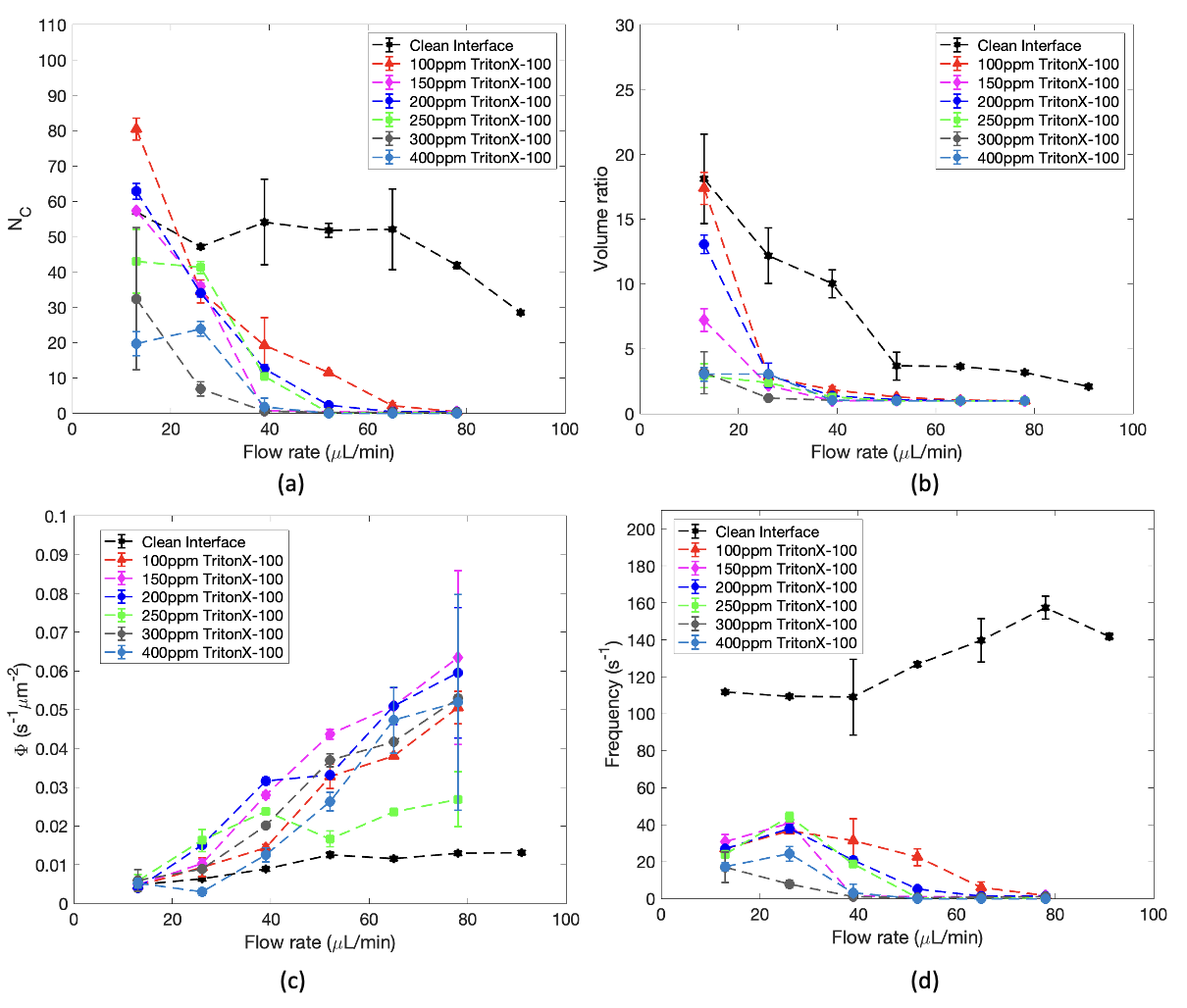} 
  \caption{Plots of (a) the number of coalescence events that happened as the droplets pass from the inlet to the outlet regions, (b) the volume ratio per droplet between inlet and outlet regions, (c) the flux of droplets entering the inlet region, and (d) the frequency of coalescence, all as a function of total flow rate for water in LMO emulsions in the presence of TritonX-100 in the water phase.}
  \label{Triton Frequency}
\end{figure}

First, the number of coalescence events that the droplets undergo as they pass from the inlet to the outlet regions is shown as a function of the total flow rate (Figure \ref{Triton Frequency}(a)). It is found that the number of coalescence events decreases with flow rate for the different concentrations. This is generally expected, since the total residence time within the chamber decreases as the flow rate increases. Second, the volume ratio per droplet is plotted with flow rate (see Figure \ref{Triton Frequency}(b)). The volume ratio, defined in equation \ref{1}, represents the ratio of the average volume per droplet in the exit region to its average volume at the inlet. This ratio is also found to decrease with flow rate. A lower number of coalescence events generally leads to a decrease in the average volume ratio per droplet. However, the relation between the number of coalescence events and the volume ratio should not always be expected to be linear, as it can vary depending on the number of droplets in the exit region. As the flow rate increases, more droplets can enter the coalescence chamber per unit of time, as estimated using the flux of droplets in Figure \ref{Triton Frequency} (c). The volume ratio represents the impact of coalescence on the overall droplet size, while the number of coalescence events represents the total coalescence occurrences experienced by the droplets. This distinction eliminates the influence of potential reductions in the volume ratio caused by droplets that do not have the opportunity to coalesce, which may occur due to unfavorable flow dynamics. Note that the flux of droplets in the presence of surfactants is significantly higher than that of the clean interface (Figure \ref{Triton Frequency} (c)). The difference in fluxes with and without surfactants is associated with droplet breakup, as will be explained later in the \enquote{Effect of Droplet Breakup on Coalescence} section.

The frequency of coalescence is then calculated using the number of coalescence events and the residence time of droplets (equations \ref{1} and \ref{2}). Figure \ref{Triton Frequency}(d) shows the results for coalescence frequency as a function of flow rate for TritonX-100, at different surfactant concentrations, as well as for the clean interface. It is observed that the frequency of coalescence in the presence of surfactants is less than that of the clean interface, and that the frequency decreases with surfactant concentration. This signifies a higher emulsion stability in the presence of surfactant, and a more stable emulsion as the surfactant concentration increases, as expected. In addition, a non-monotonic trend is observed, where the frequency initially increases between a flow rate of 13 to 26 $\mu$L/min. Beyond that flow rate, the frequency of coalescence decreases. At lower flow rates, droplets are generated and transported at a slower speed, allowing them sufficient time to interact and collide with one another. This interaction promotes coalescence. As the flow rate increases, the velocity gradient along the channel height steepens, which can increase the relative velocity between the droplets, enhancing their likelihood of collisions and subsequent coalescence. Moreover, the increase in the flow rate increases the shear forces acting on the droplets by the continuous phase, also increasing their collision frequency. However, there is a critical flow rate beyond which the frequency decreases. The drop is mainly caused by the reduced residence time, as well as collision efficiency between the droplets, as droplets start to bounce-off each other instead of coalescing. As the flow rate continues to increase, droplets are transported more rapidly through the microfluidic channels. This higher speed reduces the time available for droplets to interact and collide, thus decreasing the likelihood of coalescence. However, for the clean interface case, the critical flow rate is significantly higher than the cases with surfactants, as droplets with no surfactant require significantly less time to coalesce compared to droplets with Triton X-100, as found in the study of Chen \textit{et al.} \cite{chen_under_revision}. Therefore, even when the flow rate increases in the clean interface case, the number of coalescence events does not decrease significantly, compared to the cases with surfactants.

\begin{figure}[H]
   \includegraphics[width=\textwidth]{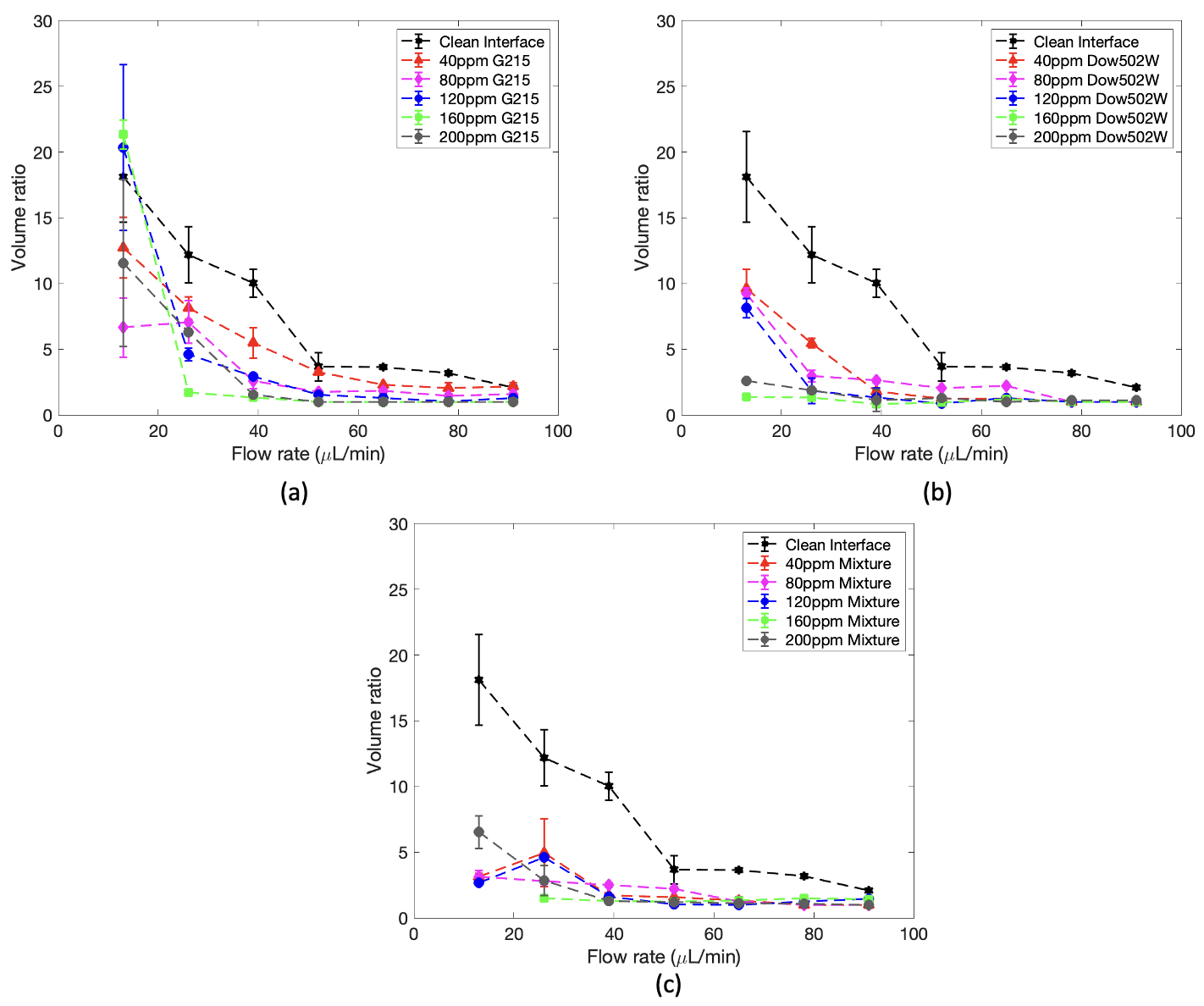} 
  \caption{Average volume ratio of the  droplets in the exit region to those in the entry region as a function of total flow rate for water in LMO emulsions in the presence of (a) G215, (b) Dow502W, and (c) 3:2 Dow502W:G215 surfactant mixture in the water phase.}
  \label{G215Dow V}
\end{figure}

In addition to Triton X-100, frequency measurements are also taken for G215 and Dow502W as well as for the surfactant mixture. The droplet volume ratio is first plotted, as shown in Figure \ref{G215Dow V} for (a) G215, (b) Dow502W, and (c) the mixture. Similar to the case of TritonX-100, the average volume ratio per droplet for  G215 and Dow502W as well as for the surfactant mixture show a general decreasing trend with flow rate. Plots of the number of coalescence events as well as droplet flux as a function of flow rate are presented in Figures SI3 and SI4, respectively.

\begin{figure}[H]
   \includegraphics[width=\textwidth]{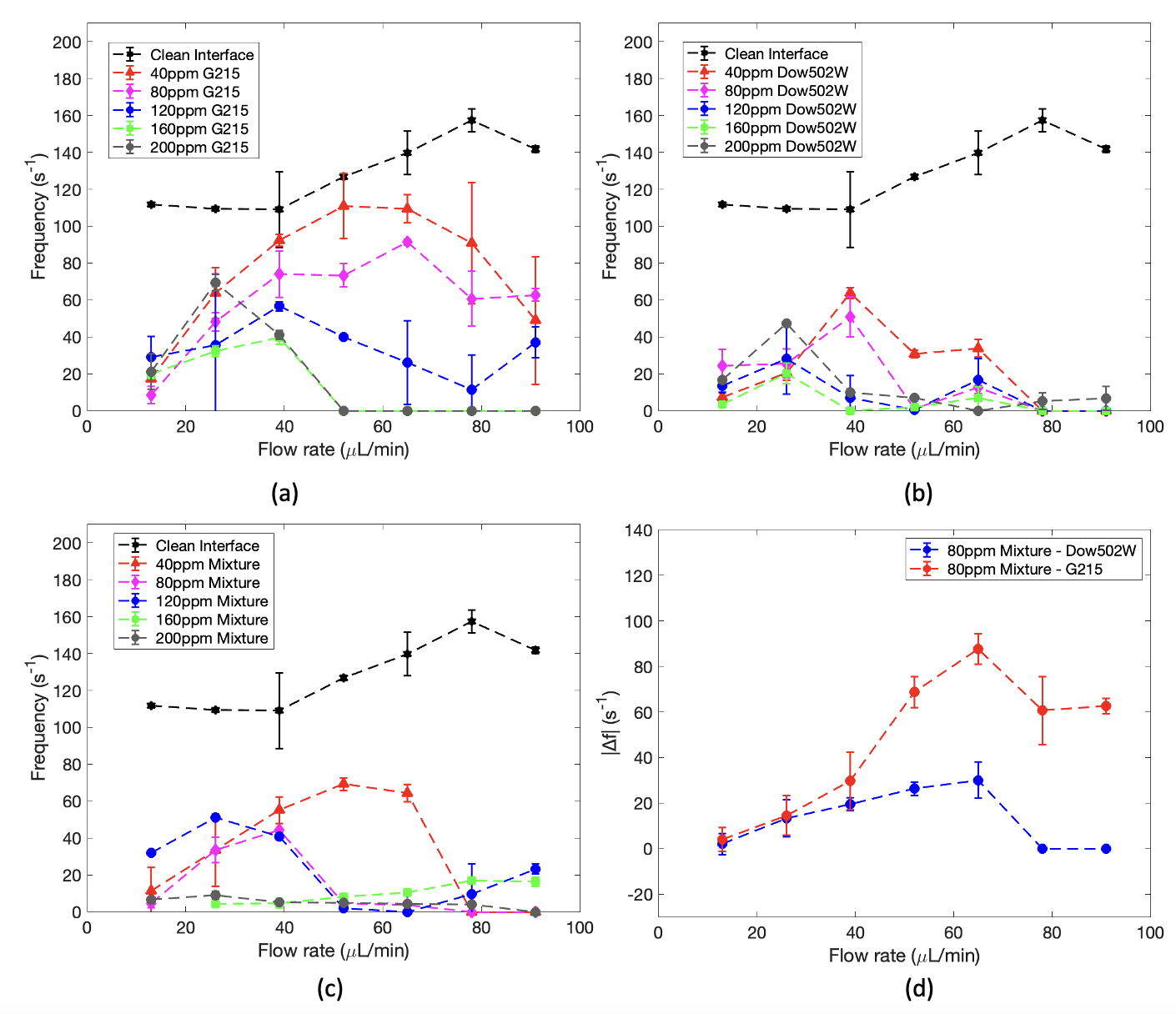} 
  \caption{Frequency of coalescence as a function of total flow rate for water in LMO emulsions in the presence of (a) G215, (b) Dow502W, and (c) 3:2 Dow502W:G215 surfactant mixture in the water phase, as well as (d) the absolute value of the difference in coalescence frequency at 80ppm between the surfactant mixture and each of Dow502W and G215. Note that the plots of absolute differences at the other concentrations are presented in Figure SI5.}
  \label{G215Dow frequency}
\end{figure}

The frequency of coalescence is then calculated, as shown in Figure \ref{G215Dow frequency} for (a) G215, (b) Dow502W, and (c) the mixture. A similar trend to that found with TritonX-100 is observed. It is also found that the frequency in the presence of surfactants is less than that of the clean interface. The frequency of coalescence in the case of Dow502W is less than that in the case of G215, suggesting that the emulsion becomes more stable in the case of Dow502W. As for the surfactant mixture case, it is shown that the frequency of coalescence is close to that of Dow502W (even slightly less at the high surfactant concentrations), suggesting a slight synergistic effect between the two surfactants. Figure \ref{G215Dow frequency} (d) shows a plot of the absolute value of the difference in coalescence frequency at 80ppm between the surfactant mixture and each of Dow502W and G215. It is apparent that the coalescence frequency of the mixture is closer to that of Dow502W than it is for the mixture. The plots showing the absolute differences for the other concentrations are provided in Figure SI5. It can be seen that the differences between the two curves are more pronounced at lower surfactant concentrations (40 and 80 ppm). 

\subsection{Non-Dimensional Analysis of Coalescence Frequency}
\subsubsection{Coalescence Frequency Normalized by Flow Rate-to-Radius Ratio}
\begin{figure}[H]
   \includegraphics[width=\textwidth]{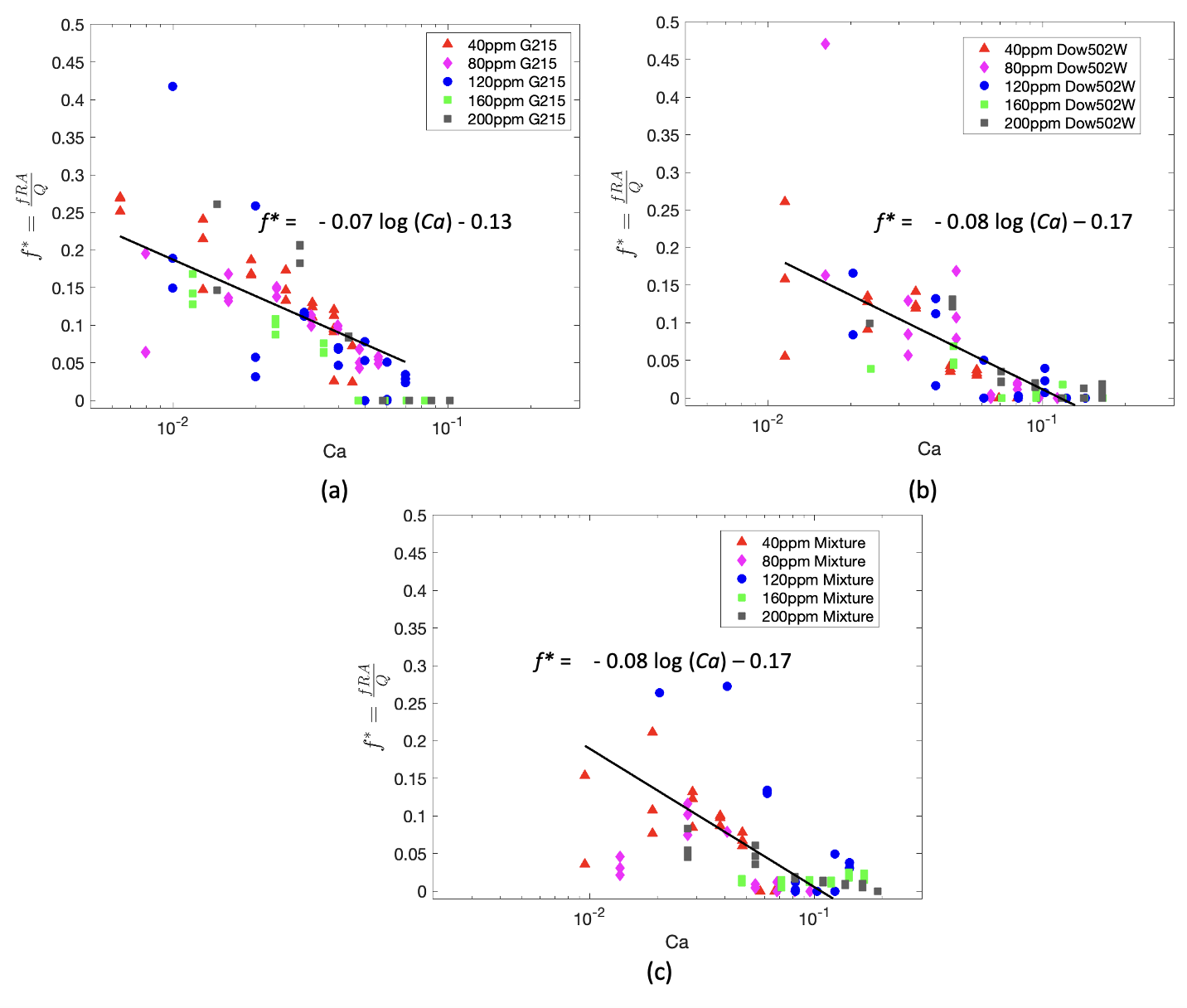} 
  \caption{Dimensionless coalescence frequency as a function of Capillary number for water in LMO emulsions in the presence of (a) G215, (b) Dow502W, and (c) 3:2 Dow502W:G215 surfactant mixture in the water phase.}
  \label{G215Dow Ca}
\end{figure}

So far, the coalescence frequency has been shown to depend on flow rate. In the study by Narayan et al. \cite{narayan2020insights,narayan2022correction}, it was observed that droplet radius influences the film drainage time between two coalescing droplets, with larger droplet sizes resulting in longer drainage times. Consequently, droplet radius can also affect coalescence frequency, potentially leading to a decrease in frequency at larger radii. To eliminate the influence of both flow rate and radius, we define a non-dimensional frequency as: $f^* = \frac{fRA}{Q}$, where $R$ is the initial droplet radius before it enters the coalescence chamber, $A$ is the cross-sectional area of the equilibration channel, and $Q$ is the total flow rate. The non-dimensional frequency is plotted as a function of Capillary number $Ca = \frac{\mu Q}{\gamma A}$, where $\mu$ is the continuous phase viscosity and $\gamma$ is the equilibrium IFT, as depicted in Figure \ref{G215Dow Ca}. The frequency curves for all concentrations collapse into a single curve, suggesting a universal behavior across different conditions. The collapse of the curves implies that the key factors governing coalescence—such as droplet size and flow dynamics—are effectively captured by the normalization, leading to a concentration-independent scaling of the coalescence frequency with $Ca$. The flow rate $Q$ is a critical factor in the coalescence of droplets. When normalizing the coalescence frequency and plotting it with $Ca$, the influence of IFT becomes less apparent when considered in conjunction with flow dynamics, and the relative difference in IFT across different concentrations may not significantly impact the normalized frequency, as the flow-induced stresses (via $Q$) dominate the behavior. In Figure \ref{G215Dow frequency}, it is noted that for a given flow rate, the increase in surfactant concentration decreases the coalescence frequency. Increasing the surfactant concentration also decreases the IFT, as illustrated in Figure \ref{Pendant}. Therefore, for a given flow rate, at a higher surfactant concentration, $Ca$ increases and the normalized frequency decreases, resulting in a shift of the data point at higher concentration to a  higher $Ca$ value on the x-axis. 

What is notable as well is that at high $Ca$, the non-dimensional coalescence frequency in the case of Dow502W and the surfactant mixture is lower than that in the case of G215. In general, at high $Ca$, even when the droplets have low contact time, it was still sufficient for coalescence in G215.

In addition, it can be seen that the non-dimensional frequency in the case of G215 shows a weaker dependence on $Ca$, followed by Dow502W, and then the mixture. This is revealed through fitting a logarithmic curve to the $f^*$ versus $Ca$ plots. The reason for the difference in $Ca$ dependence could be attributed to differences in interfacial elasticity, as well as differences in surfactant packing on the interface.

\subsubsection{Coalescence Frequency Normalized by Flux}
\begin{figure}[H]
   \includegraphics[width=\textwidth]{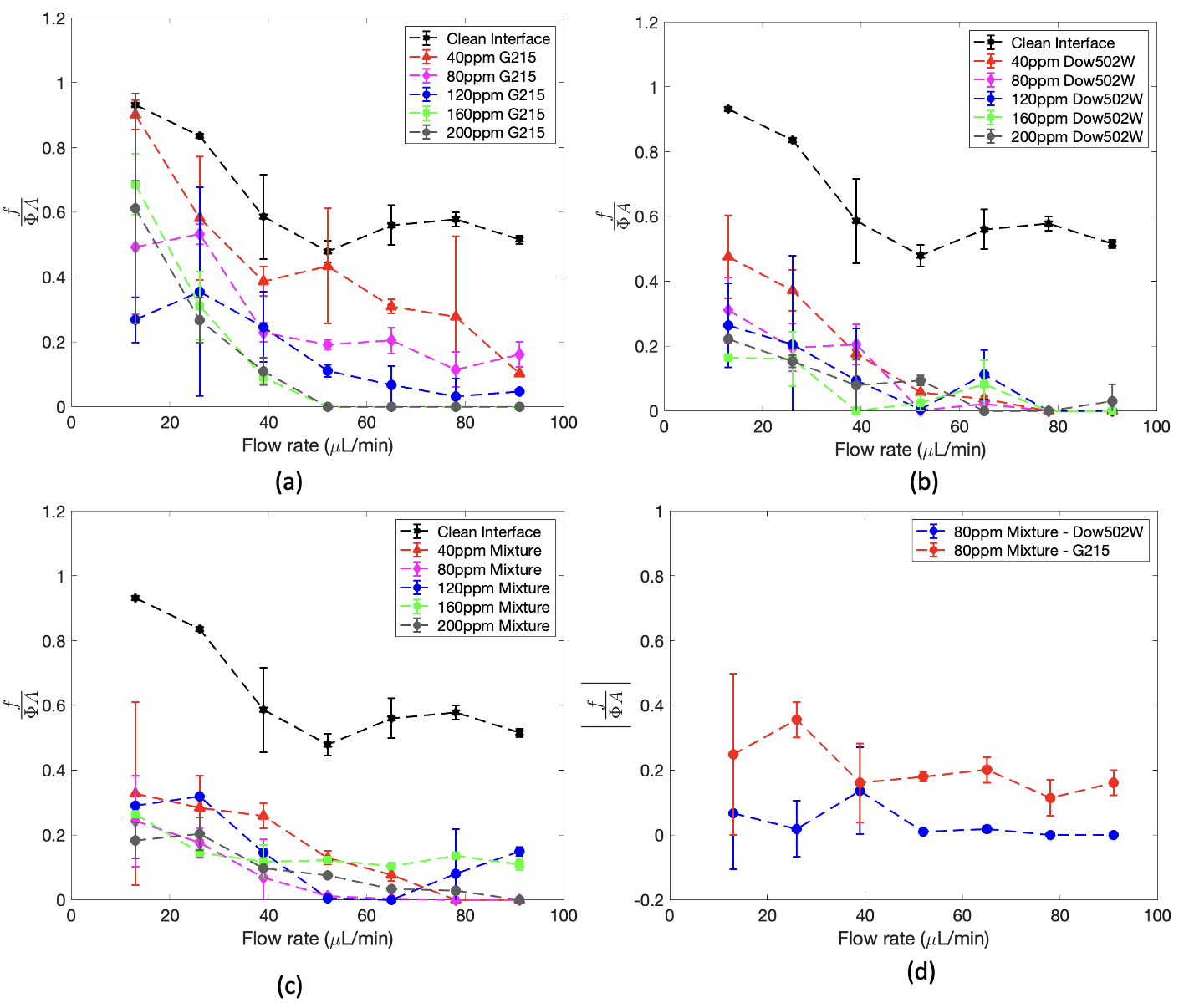} 
  \caption{Dimensionless coalescence frequency normalized by the incoming droplet flux per unit time as a function of the total flow rate for water in LMO emulsions in the presence of (a) G215, (b) Dow502W, and (c) 3:2 Dow502W:G215 surfactant mixture in the water phase. The plot in (d) shows the absolute value of the difference in non-dimensional frequency at 80ppm between the surfactant mixture and each of Dow502W and G215. Similar plots are shown for the other concentrations in Figure SI6.}
  \label{G215Dow fAflux}
\end{figure}

Previously in section \enquote{Coalescence Frequency Results}, we mentioned that the coalescence frequency is calculated based on the total number of coalescence events. This approach accounts for small droplets that reach the exit region without coalescing due to flow dynamics, which would reduce the volume ratio. However, the number of coalescence events is highly affected by the flux of incoming droplets, which has been shown to vary across concentrations, even at the same flow rate. Moreover, in some cases, the number of coalescence events increases when transitioning between flow rates at a fixed surfactant concentration, while the volume ratio decreases. An example of such events is the case of 200ppm G215 between 13 and 26 \(\mu\text{L/min}\). This could be due to the presence of more droplets at the exit region causing the increase in the total number of coalescence events, although the actual volume ratio decreases. The increase in the number of droplets at the exit region is caused by the increase in the droplet flux. To address the complexities caused by the varying flux and to eliminate the flux dependence, we normalize the coalescence frequency by the flux of droplets multiplied by the cross sectional area of the inlet channel to get the non-dimensional frequency: $\frac{f}{\phi A}$. The results are presented in Figure \ref{G215Dow fAflux}, for each of (a)G215, (b) Dow502W, and (c) the mixture system. Additionally, the absolute value of the difference in $\frac{f}{\phi A}$ between the mixture and each of G215 and Dow502W at 80ppm is plotted in Figure \ref{G215Dow fAflux} (d). Similar plots at 40, 120, 160, and 200ppm are presented in Figure SI6.

The results indicate that when the coalescence frequency is normalized by $\phi A$, which represents the number of droplets entering the inlet region per unit time, the non-dimensional frequency decreases with flow rate. This indicates that the initial increase  in frequency with flow rate in Figure \ref{G215Dow frequency} might be caused by the increase in the droplet flux causing more coalescence events to occur. In general, similar to the dimensionless frequency, the non-dimensional frequency also decreases with surfactant concentration. In addition, it is still notable that the non-dimensional frequency for the surfactant mixture and Dow502W is less than that for G215. The non-dimensional frequency in the case of the surfactant mixture is also closer to that of Dow502W, than it to that of G215, as depicted in Figure \ref{G215Dow fAflux} and Figure SI6, especially at low surfactant concentrations (40 and 80ppm).

\subsection{Effect of Droplet Breakup on Coalescence}
\begin{figure}[H]
   \includegraphics[width=\textwidth]{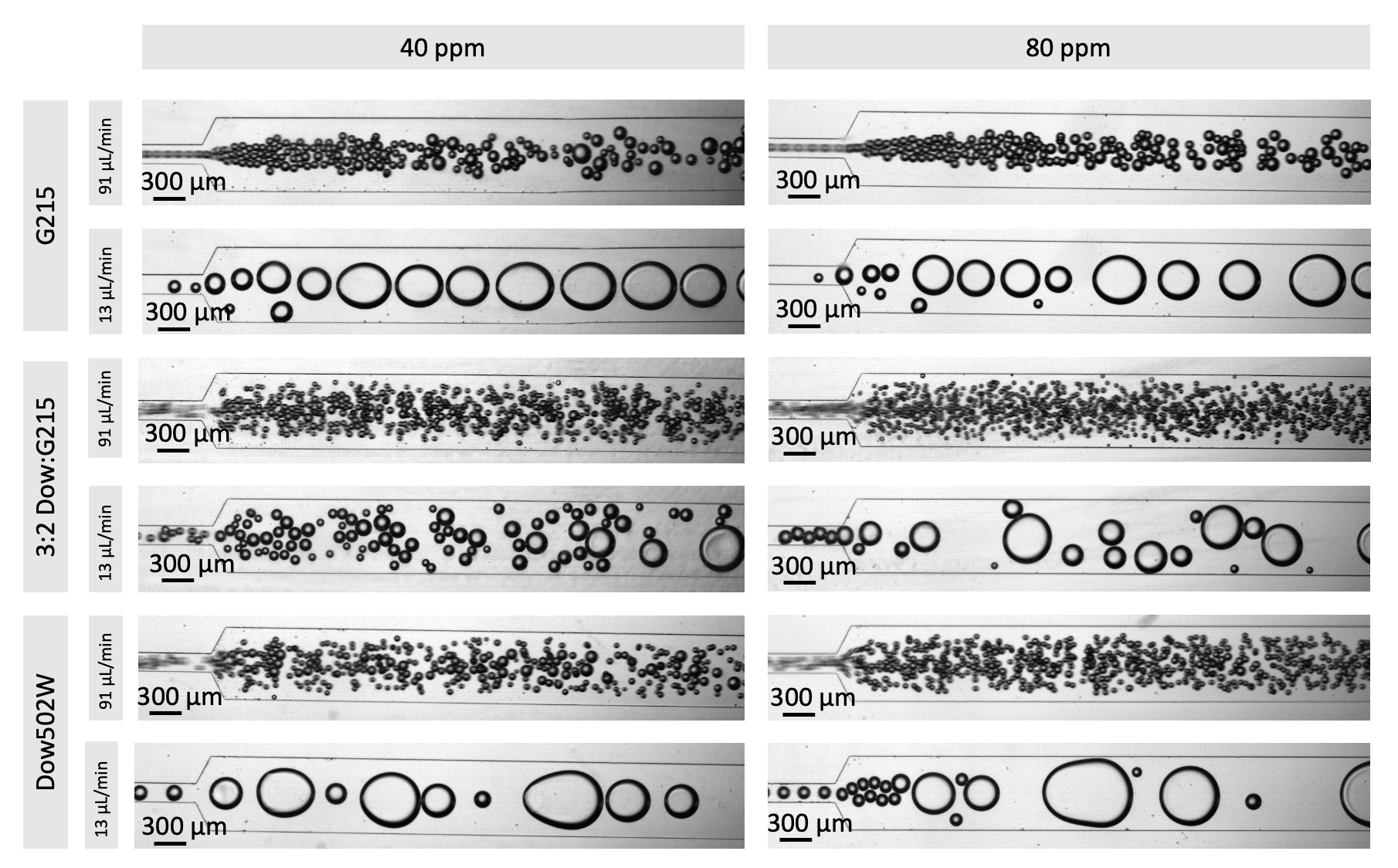} 
  \caption{Screenshot of coalescence events in the microfluidic coalescence channel illustrating droplet breakup occurring in the cases of Dow502W as well as the 3:2 Dow502W:G215 surfactant mixture, at the highest flow rate of 91 $\mu$L/min, compared to droplets not breaking up and undergoing coalescence at the lowest flow rate of 13  $\mu$L/min}
  \label{Droplet breakup}
\end{figure}

As mentioned previously, the ratio of flow rates between the continuous and dispersed phases was kept constant across experiments to maintain a similar droplet radius. However, in certain cases, especially in the cases of TritonX-100, Dow502W as well as the surfactant mixture, at high flow rates, smaller sized droplets were observed (see Figure \ref{Droplet breakup}). This was due to the droplets breaking-up in the T-junction into smaller droplets. The droplets breakup due to the lower IFT values which make it easier for the droplets to deform and pinch off. In the cases when droplets break-up, the continuous flow in the coalescence chamber carries the smaller droplets apart, reducing their chance of collision. This causes those systems to be more stable against coalescence, and thus reduce their coalescence frequency. This is observed in Figure \ref{Droplet breakup}, at the high flow rate of 91 $\mu$L/min. Two example concentrations are illustrated: 40 and 80 ppm for each of Dow502W and the surfactant mixture. At both concentrations and for both systems, little to no coalescence was observed at the high flow rate of 91 $\mu$L/min, and coalescence was observed at the low flow rate of 13 $\mu$L/min, during which no break-up occurs. Similar observations were found for TritonX-100, where droplet breakup occurs at high flow rates. However, for the case of G215, no breakup occurs at at both 13 $\mu$L/min and 91 $\mu$L/min, at a surfactant concentration of 40 and 80 ppm. However, at higher concentrations (120, 160, and 200ppm), the IFT decreases further, and droplet breakup is observed in the case of G215. G215 has a relatively higher IFT than both Dow502W and the surfactant mixture, as depicted in Figure \ref{Pendant}. This explains why breakup does not occur in G215 at low concentrations.

A similar observation was shown in the study of Grace \textit{et al.}\cite{grace1982dispersion} , where droplets breakup above a certain critical $Ca$, and a higher IFT increases the critical $Ca$ required for breakup. Droplet breakup is not observed in our experiments for the clean interface as well as G215 due to the relatively high IFT, suggesting that a higher $Ca$ is required for breakup. However, in the case of surfactants that are more effective in lowering the IFT (TritonX-100, Dow502W, and the surfactant mixture), breakup occurred at high flow rates, and thus at high $Ca$. 

\section{Conclusion}
In summary, this work suggests the use of surfactant replacements in the context of firefighting foams. Three different surfactants were tested: TritonX-100, G215, and Dow502W. In addition, a mixture of 3:2 G215:Dow502W by weight was tested. Stability effectiveness was measured by evaluating the stability of a water in LMO emulsion with the different surfactants dissolved in the water phase. Emulsion stability is evaluated using a microfluidic device used to find coalescence frequency, as a function of total flow rate. It is found that the coalescence frequency decreases with surfactant concentration, and varies non-monotonically with total flow rate. Moreover, the coalescence frequency of the surfactant mixture is found to be close to that of Dow502W, suggesting a synergistic behavior between surfactants. In addition, IFT measurements are done for water in LMO with G215, Dow502W, and the surfactant mixture. It was also found that the IFT in the case of the surfactant mixture is close to that in the case of Dow502W, suggesting that the faster surfactant dominates. 

\begin{acknowledgement}
This material is based upon initial project work supported by the Humphreys Engineer Center support activity under contract No. W912HQ20C0041, corresponding to The Department of Defense Strategic Environmental Research and Development Program (SERDP) project WP19-1407.  Additional project work was then supported by The US Army Corps of Engineers and SERDP under contract No. W912HQ24C0067, corresponding to SERDP project WP24-4147. The authors acknowledge the helpful discussions with Dr. Yun Chen and Zak Kujala. The authors also thank professor Michael Manno, and Dr. Wieslaw Suszynski for the use of the pendant drop apparatus in the Coatings Process Fundamentals Lab in the Department of Chemical Engineering at the University of Minnesota. Portions of this work were conducted in the Minnesota Nano Center, which is supported by the National Science Foundation through the National Nanotechnology Coordinated Infrastructure (NNCI) under Award Number ECCS-2025124. None of the authors have recent, present, or anticipated financial gain from this work. Figure 1 is created using BioRender.com. 
\end{acknowledgement}

\begin{suppinfo}
Dynamic IFT data from pendant drop, dynamic IFT data in the presence of G215 at 40 ppm from microfluidics, number of coalescence events as a function of flow rate for each of G215, Dow502W, and the surfactant mixture, flux of droplets as a function of flow rate for each of G215, Dow502W, and the surfactant mixture, absolute value of difference in coalescence frequency between mixture and each of G215 and Dow502W, absolute value of difference in non-dimensional frequency between mixture and each of G215 and Dow502W

\end{suppinfo}

\bibliography{achemso-demo}

\providecommand{\latin}[1]{#1}
\makeatletter
\providecommand{\doi}
  {\begingroup\let\do\@makeother\dospecials
  \catcode`\{=1 \catcode`\}=2 \doi@aux}
\providecommand{\doi@aux}[1]{\endgroup\texttt{#1}}
\makeatother
\providecommand*\mcitethebibliography{\thebibliography}
\csname @ifundefined\endcsname{endmcitethebibliography}  {\let\endmcitethebibliography\endthebibliography}{}
\begin{mcitethebibliography}{44}
\providecommand*\natexlab[1]{#1}
\providecommand*\mciteSetBstSublistMode[1]{}
\providecommand*\mciteSetBstMaxWidthForm[2]{}
\providecommand*\mciteBstWouldAddEndPuncttrue
  {\def\EndOfBibitem{\unskip.}}
\providecommand*\mciteBstWouldAddEndPunctfalse
  {\let\EndOfBibitem\relax}
\providecommand*\mciteSetBstMidEndSepPunct[3]{}
\providecommand*\mciteSetBstSublistLabelBeginEnd[3]{}
\providecommand*\EndOfBibitem{}
\mciteSetBstSublistMode{f}
\mciteSetBstMaxWidthForm{subitem}{(\alph{mcitesubitemcount})}
\mciteSetBstSublistLabelBeginEnd
  {\mcitemaxwidthsubitemform\space}
  {\relax}
  {\relax}

\bibitem[Prud'Homme and Khan(2017)Prud'Homme, and Khan]{prud2017foams}
Prud'Homme,~R.~K.; Khan,~S.~A. \emph{Foams: theory: measurements: applications}; Routledge, 2017\relax
\mciteBstWouldAddEndPuncttrue
\mciteSetBstMidEndSepPunct{\mcitedefaultmidpunct}
{\mcitedefaultendpunct}{\mcitedefaultseppunct}\relax
\EndOfBibitem
\bibitem[Brice{\~n}o-Ahumada \latin{et~al.}(2021)Brice{\~n}o-Ahumada, Soltero-Mart{\'\i}nez, and Castillo]{briceno2021aqueous}
Brice{\~n}o-Ahumada,~Z.; Soltero-Mart{\'\i}nez,~J.; Castillo,~R. Aqueous foams and emulsions stabilized by mixtures of silica nanoparticles and surfactants: A state-of-the-art review. \emph{Chemical Engineering Journal Advances} \textbf{2021}, \emph{7}, 100116\relax
\mciteBstWouldAddEndPuncttrue
\mciteSetBstMidEndSepPunct{\mcitedefaultmidpunct}
{\mcitedefaultendpunct}{\mcitedefaultseppunct}\relax
\EndOfBibitem
\bibitem[Langevin(2023)]{langevin2023recent}
Langevin,~D. Recent advances on emulsion and foam stability. \emph{Langmuir} \textbf{2023}, \emph{39}, 3821--3828\relax
\mciteBstWouldAddEndPuncttrue
\mciteSetBstMidEndSepPunct{\mcitedefaultmidpunct}
{\mcitedefaultendpunct}{\mcitedefaultseppunct}\relax
\EndOfBibitem
\bibitem[Matusiak \latin{et~al.}(2020)Matusiak, Grzadka, Kowalczuk, Pietruszka, and Godlewski]{matusiak2020influence}
Matusiak,~J.; Grzadka,~E.; Kowalczuk,~A.; Pietruszka,~R.; Godlewski,~M. The influence of hydrocarbon, fluorinated and silicone surfactants on the adsorption, stability and electrokinetic properties of the $\kappa$-carrageenan/alumina system. \emph{Journal of Molecular Liquids} \textbf{2020}, \emph{314}, 113669\relax
\mciteBstWouldAddEndPuncttrue
\mciteSetBstMidEndSepPunct{\mcitedefaultmidpunct}
{\mcitedefaultendpunct}{\mcitedefaultseppunct}\relax
\EndOfBibitem
\bibitem[Zhao \latin{et~al.}(2024)Zhao, Cheng, Pu, Su, Wang, Cao, and Liu]{zhao2024research}
Zhao,~W.; Cheng,~Y.; Pu,~J.; Su,~L.; Wang,~N.; Cao,~Y.; Liu,~L. Research Progress in Structure Synthesis, Properties, and Applications of Small-Molecule Silicone Surfactants. \emph{Topics in Current Chemistry} \textbf{2024}, \emph{382}, 11\relax
\mciteBstWouldAddEndPuncttrue
\mciteSetBstMidEndSepPunct{\mcitedefaultmidpunct}
{\mcitedefaultendpunct}{\mcitedefaultseppunct}\relax
\EndOfBibitem
\bibitem[Hunter \latin{et~al.}(2008)Hunter, Pugh, Franks, and Jameson]{hunter2008role}
Hunter,~T.~N.; Pugh,~R.~J.; Franks,~G.~V.; Jameson,~G.~J. The role of particles in stabilising foams and emulsions. \emph{Advances in Colloid and Interface Science} \textbf{2008}, \emph{137}, 57--81\relax
\mciteBstWouldAddEndPuncttrue
\mciteSetBstMidEndSepPunct{\mcitedefaultmidpunct}
{\mcitedefaultendpunct}{\mcitedefaultseppunct}\relax
\EndOfBibitem
\bibitem[Krafft(2001)]{krafft2001fluorocarbons}
Krafft,~M.~P. Fluorocarbons and fluorinated amphiphiles in drug delivery and biomedical research. \emph{Advanced Drug Delivery Reviews} \textbf{2001}, \emph{47}, 209--228\relax
\mciteBstWouldAddEndPuncttrue
\mciteSetBstMidEndSepPunct{\mcitedefaultmidpunct}
{\mcitedefaultendpunct}{\mcitedefaultseppunct}\relax
\EndOfBibitem
\bibitem[Czajka \latin{et~al.}(2015)Czajka, Hazell, and Eastoe]{czajka2015surfactants}
Czajka,~A.; Hazell,~G.; Eastoe,~J. Surfactants at the design limit. \emph{Langmuir} \textbf{2015}, \emph{31}, 8205--8217\relax
\mciteBstWouldAddEndPuncttrue
\mciteSetBstMidEndSepPunct{\mcitedefaultmidpunct}
{\mcitedefaultendpunct}{\mcitedefaultseppunct}\relax
\EndOfBibitem
\bibitem[Hussain \latin{et~al.}(2022)Hussain, Adewunmi, Mahboob, Murtaza, Zhou, and Kamal]{hussain2022fluorinated}
Hussain,~S. M.~S.; Adewunmi,~A.~A.; Mahboob,~A.; Murtaza,~M.; Zhou,~X.; Kamal,~M.~S. Fluorinated surfactants: A review on recent progress on synthesis and oilfield applications. \emph{Advances in Colloid and Interface Science} \textbf{2022}, \emph{303}, 102634\relax
\mciteBstWouldAddEndPuncttrue
\mciteSetBstMidEndSepPunct{\mcitedefaultmidpunct}
{\mcitedefaultendpunct}{\mcitedefaultseppunct}\relax
\EndOfBibitem
\bibitem[Pabon and Corpart(2002)Pabon, and Corpart]{pabon2002fluorinated}
Pabon,~M.; Corpart,~J. Fluorinated surfactants: synthesis, properties, effluent treatment. \emph{Journal of Fluorine Chemistry} \textbf{2002}, \emph{114}, 149--156\relax
\mciteBstWouldAddEndPuncttrue
\mciteSetBstMidEndSepPunct{\mcitedefaultmidpunct}
{\mcitedefaultendpunct}{\mcitedefaultseppunct}\relax
\EndOfBibitem
\bibitem[Porter(2013)]{porter2013handbook}
Porter,~M.~R. \emph{Handbook of surfactants}; Springer, 2013\relax
\mciteBstWouldAddEndPuncttrue
\mciteSetBstMidEndSepPunct{\mcitedefaultmidpunct}
{\mcitedefaultendpunct}{\mcitedefaultseppunct}\relax
\EndOfBibitem
\bibitem[Bell \latin{et~al.}(2021)Bell, De~Guise, McCutcheon, Lei, Levin, Li, Rusling, Lawrence, Cavallari, O'Connell, \latin{et~al.} others]{bell2021exposure}
Bell,~E.~M.; De~Guise,~S.; McCutcheon,~J.~R.; Lei,~Y.; Levin,~M.; Li,~B.; Rusling,~J.~F.; Lawrence,~D.~A.; Cavallari,~J.~M.; O'Connell,~C.; others Exposure, health effects, sensing, and remediation of the emerging PFAS contaminants--Scientific challenges and potential research directions. \emph{Science of the Total Environment} \textbf{2021}, \emph{780}, 146399\relax
\mciteBstWouldAddEndPuncttrue
\mciteSetBstMidEndSepPunct{\mcitedefaultmidpunct}
{\mcitedefaultendpunct}{\mcitedefaultseppunct}\relax
\EndOfBibitem
\bibitem[Sunderland \latin{et~al.}(2019)Sunderland, Hu, Dassuncao, Tokranov, Wagner, and Allen]{sunderland2019review}
Sunderland,~E.~M.; Hu,~X.~C.; Dassuncao,~C.; Tokranov,~A.~K.; Wagner,~C.~C.; Allen,~J.~G. A review of the pathways of human exposure to poly-and perfluoroalkyl substances (PFASs) and present understanding of health effects. \emph{Journal of Exposure Science \& Environmental Epidemiology} \textbf{2019}, \emph{29}, 131--147\relax
\mciteBstWouldAddEndPuncttrue
\mciteSetBstMidEndSepPunct{\mcitedefaultmidpunct}
{\mcitedefaultendpunct}{\mcitedefaultseppunct}\relax
\EndOfBibitem
\bibitem[Pelch \latin{et~al.}(2019)Pelch, Reade, Wolffe, and Kwiatkowski]{pelch2019pfas}
Pelch,~K.~E.; Reade,~A.; Wolffe,~T.~A.; Kwiatkowski,~C.~F. PFAS health effects database: Protocol for a systematic evidence map. \emph{Environment International} \textbf{2019}, \emph{130}, 104851\relax
\mciteBstWouldAddEndPuncttrue
\mciteSetBstMidEndSepPunct{\mcitedefaultmidpunct}
{\mcitedefaultendpunct}{\mcitedefaultseppunct}\relax
\EndOfBibitem
\bibitem[Hetzer \latin{et~al.}(2014)Hetzer, K{\"u}mmerlen, Wirz, and Blunk]{hetzer2014fire}
Hetzer,~R.; K{\"u}mmerlen,~F.; Wirz,~K.; Blunk,~D. Fire testing a new fluorine-free AFFF based on a novel class of environmentally sound high performance siloxane surfactants. \emph{Fire Safety Science} \textbf{2014}, \emph{11}, 1261--1270\relax
\mciteBstWouldAddEndPuncttrue
\mciteSetBstMidEndSepPunct{\mcitedefaultmidpunct}
{\mcitedefaultendpunct}{\mcitedefaultseppunct}\relax
\EndOfBibitem
\bibitem[Kaller \latin{et~al.}(2023)Kaller, Van~Bortel, Engels, Thierens, and Fachinger]{kaller2023evaluation}
Kaller,~M.; Van~Bortel,~G.; Engels,~T.; Thierens,~R.; Fachinger,~J. An evaluation of the firefighting performance of alcohol-resistant aqueous film forming foams (AFFF-AR) and alcohol-resistant fluorine-free foams (FFF-AR) in the past two decades. \emph{Fire Technology} \textbf{2023}, \emph{59}, 429--452\relax
\mciteBstWouldAddEndPuncttrue
\mciteSetBstMidEndSepPunct{\mcitedefaultmidpunct}
{\mcitedefaultendpunct}{\mcitedefaultseppunct}\relax
\EndOfBibitem
\bibitem[Islam and Lattimer(2024)Islam, and Lattimer]{islam2024understanding}
Islam,~R.; Lattimer,~B.~Y. Understanding the Impact of Fuel on Surfactant Microstructure of Firefighting Foam. \emph{Fire Technology} \textbf{2024}, 1--29\relax
\mciteBstWouldAddEndPuncttrue
\mciteSetBstMidEndSepPunct{\mcitedefaultmidpunct}
{\mcitedefaultendpunct}{\mcitedefaultseppunct}\relax
\EndOfBibitem
\bibitem[Benali \latin{et~al.}(2021)Benali, S{\'a}, Pinho, Fernandes, and Pereira]{benali2021understanding}
Benali,~A.; S{\'a},~A.~C.; Pinho,~J.; Fernandes,~P.~M.; Pereira,~J.~M. Understanding the impact of different landscape-level fuel management strategies on wildfire hazard in central Portugal. \emph{Forests} \textbf{2021}, \emph{12}, 522\relax
\mciteBstWouldAddEndPuncttrue
\mciteSetBstMidEndSepPunct{\mcitedefaultmidpunct}
{\mcitedefaultendpunct}{\mcitedefaultseppunct}\relax
\EndOfBibitem
\bibitem[Hinnant \latin{et~al.}(2018)Hinnant, Giles, Snow, Farley, Fleming, and Ananth]{hinnant2018analytically}
Hinnant,~K.~M.; Giles,~S.~L.; Snow,~A.~W.; Farley,~J.~P.; Fleming,~J.~W.; Ananth,~R. An analytically defined fire-suppressing foam formulation for evaluation of fluorosurfactant replacement. \emph{Journal of Surfactants and Detergents} \textbf{2018}, \emph{21}, 711--722\relax
\mciteBstWouldAddEndPuncttrue
\mciteSetBstMidEndSepPunct{\mcitedefaultmidpunct}
{\mcitedefaultendpunct}{\mcitedefaultseppunct}\relax
\EndOfBibitem
\bibitem[Ananth \latin{et~al.}(2019)Ananth, Snow, Hinnant, Giles, and Farley]{ananth2019synergisms}
Ananth,~R.; Snow,~A.~W.; Hinnant,~K.~M.; Giles,~S.~L.; Farley,~J.~P. Synergisms between siloxane-polyoxyethylene and alkyl polyglycoside surfactants in foam stability and pool fire extinction. \emph{Colloids and Surfaces A: Physicochemical and Engineering Aspects} \textbf{2019}, \emph{579}, 123686\relax
\mciteBstWouldAddEndPuncttrue
\mciteSetBstMidEndSepPunct{\mcitedefaultmidpunct}
{\mcitedefaultendpunct}{\mcitedefaultseppunct}\relax
\EndOfBibitem
\bibitem[Moody and Field(2000)Moody, and Field]{moody2000perfluorinated}
Moody,~C.~A.; Field,~J.~A. Perfluorinated surfactants and the environmental implications of their use in fire-fighting foams. \emph{Environmental science \& Technology} \textbf{2000}, \emph{34}, 3864--3870\relax
\mciteBstWouldAddEndPuncttrue
\mciteSetBstMidEndSepPunct{\mcitedefaultmidpunct}
{\mcitedefaultendpunct}{\mcitedefaultseppunct}\relax
\EndOfBibitem
\bibitem[Buck \latin{et~al.}(2011)Buck, Franklin, Berger, Conder, Cousins, De~Voogt, Jensen, Kannan, Mabury, and van Leeuwen]{buck2011perfluoroalkyl}
Buck,~R.~C.; Franklin,~J.; Berger,~U.; Conder,~J.~M.; Cousins,~I.~T.; De~Voogt,~P.; Jensen,~A.~A.; Kannan,~K.; Mabury,~S.~A.; van Leeuwen,~S.~P. Perfluoroalkyl and polyfluoroalkyl substances in the environment: terminology, classification, and origins. \emph{Integrated Environmental Assessment and Management} \textbf{2011}, \emph{7}, 513--541\relax
\mciteBstWouldAddEndPuncttrue
\mciteSetBstMidEndSepPunct{\mcitedefaultmidpunct}
{\mcitedefaultendpunct}{\mcitedefaultseppunct}\relax
\EndOfBibitem
\bibitem[Bera \latin{et~al.}(2013)Bera, Ojha, and Mandal]{bera2013synergistic}
Bera,~A.; Ojha,~K.; Mandal,~A. Synergistic effect of mixed surfactant systems on foam behavior and surface tension. \emph{Journal of Surfactants and Detergents} \textbf{2013}, \emph{16}, 621--630\relax
\mciteBstWouldAddEndPuncttrue
\mciteSetBstMidEndSepPunct{\mcitedefaultmidpunct}
{\mcitedefaultendpunct}{\mcitedefaultseppunct}\relax
\EndOfBibitem
\bibitem[del Burgo \latin{et~al.}(2007)del Burgo, Aicart, and Junquera]{del2007mixed}
del Burgo,~P.; Aicart,~E.; Junquera,~E. Mixed vesicles and mixed micelles of the cationic--cationic surfactant system: didecyldimethylammonium bromide/dodecylethyldimethylammonium bromide/water. \emph{Colloids and Surfaces A: Physicochemical and Engineering Aspects} \textbf{2007}, \emph{292}, 165--172\relax
\mciteBstWouldAddEndPuncttrue
\mciteSetBstMidEndSepPunct{\mcitedefaultmidpunct}
{\mcitedefaultendpunct}{\mcitedefaultseppunct}\relax
\EndOfBibitem
\bibitem[Kume \latin{et~al.}(2008)Kume, Gallotti, and Nunes]{kume2008review}
Kume,~G.; Gallotti,~M.; Nunes,~G. Review on anionic/cationic surfactant mixtures. \emph{Journal of Surfactants and Detergents} \textbf{2008}, \emph{11}, 1--11\relax
\mciteBstWouldAddEndPuncttrue
\mciteSetBstMidEndSepPunct{\mcitedefaultmidpunct}
{\mcitedefaultendpunct}{\mcitedefaultseppunct}\relax
\EndOfBibitem
\bibitem[Krebs \latin{et~al.}(2012)Krebs, Schroen, and Boom]{krebs2012microfluidic}
Krebs,~T.; Schroen,~K.; Boom,~R. A microfluidic method to study demulsification kinetics. \emph{Lab on a Chip} \textbf{2012}, \emph{12}, 1060--1070\relax
\mciteBstWouldAddEndPuncttrue
\mciteSetBstMidEndSepPunct{\mcitedefaultmidpunct}
{\mcitedefaultendpunct}{\mcitedefaultseppunct}\relax
\EndOfBibitem
\bibitem[Baret \latin{et~al.}(2009)Baret, Kleinschmidt, El~Harrak, and Griffiths]{baret2009kinetic}
Baret,~J.-C.; Kleinschmidt,~F.; El~Harrak,~A.; Griffiths,~A.~D. Kinetic aspects of emulsion stabilization by surfactants: a microfluidic analysis. \emph{Langmuir} \textbf{2009}, \emph{25}, 6088--6093\relax
\mciteBstWouldAddEndPuncttrue
\mciteSetBstMidEndSepPunct{\mcitedefaultmidpunct}
{\mcitedefaultendpunct}{\mcitedefaultseppunct}\relax
\EndOfBibitem
\bibitem[Bremond \latin{et~al.}(2008)Bremond, Thiam, and Bibette]{bremond2008decompressing}
Bremond,~N.; Thiam,~A.~R.; Bibette,~J. Decompressing emulsion droplets favors coalescence. \emph{Physical Review Letters} \textbf{2008}, \emph{100}, 024501\relax
\mciteBstWouldAddEndPuncttrue
\mciteSetBstMidEndSepPunct{\mcitedefaultmidpunct}
{\mcitedefaultendpunct}{\mcitedefaultseppunct}\relax
\EndOfBibitem
\bibitem[Krebs \latin{et~al.}(2013)Krebs, Schro{\"e}n, and Boom]{krebs2013coalescence}
Krebs,~T.; Schro{\"e}n,~C.; Boom,~R. Coalescence kinetics of oil-in-water emulsions studied with microfluidics. \emph{Fuel} \textbf{2013}, \emph{106}, 327--334\relax
\mciteBstWouldAddEndPuncttrue
\mciteSetBstMidEndSepPunct{\mcitedefaultmidpunct}
{\mcitedefaultendpunct}{\mcitedefaultseppunct}\relax
\EndOfBibitem
\bibitem[Tan \latin{et~al.}(2004)Tan, Fisher, Lee, Cristini, and Lee]{tan2004design}
Tan,~Y.-C.; Fisher,~J.~S.; Lee,~A.~I.; Cristini,~V.; Lee,~A.~P. Design of microfluidic channel geometries for the control of droplet volume, chemical concentration, and sorting. \emph{Lab on a Chip} \textbf{2004}, \emph{4}, 292--298\relax
\mciteBstWouldAddEndPuncttrue
\mciteSetBstMidEndSepPunct{\mcitedefaultmidpunct}
{\mcitedefaultendpunct}{\mcitedefaultseppunct}\relax
\EndOfBibitem
\bibitem[Narayan \latin{et~al.}(2020)Narayan, Makhnenko, Moravec, Hauser, Dallas, and Dutcher]{narayan2020insights}
Narayan,~S.; Makhnenko,~I.; Moravec,~D.~B.; Hauser,~B.~G.; Dallas,~A.~J.; Dutcher,~C.~S. Insights into the microscale coalescence behavior of surfactant-stabilized droplets using a microfluidic hydrodynamic trap. \emph{Langmuir} \textbf{2020}, \emph{36}, 9827--9842\relax
\mciteBstWouldAddEndPuncttrue
\mciteSetBstMidEndSepPunct{\mcitedefaultmidpunct}
{\mcitedefaultendpunct}{\mcitedefaultseppunct}\relax
\EndOfBibitem
\bibitem[Narayan \latin{et~al.}(2022)Narayan, Makhnenko, Moravec, Hauser, Dallas, and Dutcher]{narayan2022correction}
Narayan,~S.; Makhnenko,~I.; Moravec,~D.~B.; Hauser,~B.~G.; Dallas,~A.~J.; Dutcher,~C.~S. Correction to “insights into the microscale coalescence behavior of surfactant-stabilized droplets using a microfluidic hydrodynamic trap”. \emph{Langmuir} \textbf{2022}, \emph{38}, 2749--2750\relax
\mciteBstWouldAddEndPuncttrue
\mciteSetBstMidEndSepPunct{\mcitedefaultmidpunct}
{\mcitedefaultendpunct}{\mcitedefaultseppunct}\relax
\EndOfBibitem
\bibitem[Bachnak \latin{et~al.}(2024)Bachnak, Narayan, Moravec, Hauser, Dallas, and Dutcher]{bachnak2024influence}
Bachnak,~R.; Narayan,~S.; Moravec,~D.~B.; Hauser,~B.~G.; Dallas,~A.~J.; Dutcher,~C.~S. Influence of Aqueous Phase Salt and Oil Phase Surfactants and Viscosity on the Dynamic Interfacial Tension and Coalescence Timescales. \emph{The Journal of Physical Chemistry B} \textbf{2024}, \emph{128}, 10986--10998\relax
\mciteBstWouldAddEndPuncttrue
\mciteSetBstMidEndSepPunct{\mcitedefaultmidpunct}
{\mcitedefaultendpunct}{\mcitedefaultseppunct}\relax
\EndOfBibitem
\bibitem[Bachnak \latin{et~al.}(2023)Bachnak, Panigrahi, Moravec, Hauser, Dallas, and Dutcher]{bachnak2023effect}
Bachnak,~R.; Panigrahi,~C.; Moravec,~D.~B.; Hauser,~B.~G.; Dallas,~A.~J.; Dutcher,~C.~S. The Effect of Surface Interactions on the Coalescence of Water Droplets in Fuel. \emph{Energy \& Fuels} \textbf{2023}, \emph{37}, 15956--15966\relax
\mciteBstWouldAddEndPuncttrue
\mciteSetBstMidEndSepPunct{\mcitedefaultmidpunct}
{\mcitedefaultendpunct}{\mcitedefaultseppunct}\relax
\EndOfBibitem
\bibitem[Dudek \latin{et~al.}(2020)Dudek, Fernandes, Her{\o}, and {\O}ye]{dudek2020microfluidic}
Dudek,~M.; Fernandes,~D.; Her{\o},~E.~H.; {\O}ye,~G. Microfluidic method for determining drop-drop coalescence and contact times in flow. \emph{Colloids and Surfaces A: Physicochemical and Engineering Aspects} \textbf{2020}, \emph{586}, 124265\relax
\mciteBstWouldAddEndPuncttrue
\mciteSetBstMidEndSepPunct{\mcitedefaultmidpunct}
{\mcitedefaultendpunct}{\mcitedefaultseppunct}\relax
\EndOfBibitem
\bibitem[Chen \latin{et~al.}(2025)Chen, Bahadori, and Dutcher]{chen_under_revision}
Chen,~Y.; Bahadori,~N.; Dutcher,~C.~S. Under revision in Soft Matter\relax
\mciteBstWouldAddEndPuncttrue
\mciteSetBstMidEndSepPunct{\mcitedefaultmidpunct}
{\mcitedefaultendpunct}{\mcitedefaultseppunct}\relax
\EndOfBibitem
\bibitem[Basset \latin{et~al.}(1984)Basset, Hermant, and Martin]{basset1984oil}
Basset,~D.; Hermant,~M.; Martin,~J. Oil-soluble fluorinated compounds as antiwear and antifriction additives. \emph{ASLE transactions} \textbf{1984}, \emph{27}, 380--388\relax
\mciteBstWouldAddEndPuncttrue
\mciteSetBstMidEndSepPunct{\mcitedefaultmidpunct}
{\mcitedefaultendpunct}{\mcitedefaultseppunct}\relax
\EndOfBibitem
\bibitem[Hu \latin{et~al.}(2017)Hu, Jing, An, Ming, Bian, and Chen]{hu2017tribological}
Hu,~M.; Jing,~L.; An,~Q.; Ming,~W.; Bian,~C.; Chen,~M. Tribological properties and milling performance of HSS-Co-E tools with fluorinated surfactants-based coatings against Ti--6Al--4V. \emph{Wear} \textbf{2017}, \emph{376}, 134--142\relax
\mciteBstWouldAddEndPuncttrue
\mciteSetBstMidEndSepPunct{\mcitedefaultmidpunct}
{\mcitedefaultendpunct}{\mcitedefaultseppunct}\relax
\EndOfBibitem
\bibitem[Ananth \latin{et~al.}(2023)Ananth, Hinnant, Snow, Bunton, Karwoski~JR, Farley, Davis, and Moore]{ananth2023development}
Ananth,~R.; Hinnant,~P. K.~M.; Snow,~A.; Bunton,~C.~M.; Karwoski~JR,~S.; Farley,~J.~P.; Davis,~M.; Moore,~D. Development of Zwitterionic Additives to Aqueous Foams to Enhance Suppression of Aromatic and Aliphatic Fuel Pool-Fires. \emph{NRL Report} \textbf{2023}, \relax
\mciteBstWouldAddEndPunctfalse
\mciteSetBstMidEndSepPunct{\mcitedefaultmidpunct}
{}{\mcitedefaultseppunct}\relax
\EndOfBibitem
\bibitem[Berry \latin{et~al.}(2015)Berry, Neeson, Dagastine, Chan, and Tabor]{berry2015measurement}
Berry,~J.~D.; Neeson,~M.~J.; Dagastine,~R.~R.; Chan,~D.~Y.; Tabor,~R.~F. Measurement of surface and interfacial tension using pendant drop tensiometry. \emph{Journal of Colloid and Interface Science} \textbf{2015}, \emph{454}, 226--237\relax
\mciteBstWouldAddEndPuncttrue
\mciteSetBstMidEndSepPunct{\mcitedefaultmidpunct}
{\mcitedefaultendpunct}{\mcitedefaultseppunct}\relax
\EndOfBibitem
\bibitem[Narayan \latin{et~al.}(2018)Narayan, Moravec, Hauser, Dallas, and Dutcher]{narayan2018removing}
Narayan,~S.; Moravec,~D.~B.; Hauser,~B.~G.; Dallas,~A.~J.; Dutcher,~C.~S. Removing water from diesel fuel: understanding the impact of droplet size on dynamic interfacial tension of water-in-fuel emulsions. \emph{Energy \& Fuels} \textbf{2018}, \emph{32}, 7326--7337\relax
\mciteBstWouldAddEndPuncttrue
\mciteSetBstMidEndSepPunct{\mcitedefaultmidpunct}
{\mcitedefaultendpunct}{\mcitedefaultseppunct}\relax
\EndOfBibitem
\bibitem[Gunjan \latin{et~al.}(2021)Gunjan, Kumar, and Raj]{gunjan2021cloaked}
Gunjan,~M.~R.; Kumar,~A.; Raj,~R. Cloaked droplets on lubricant-infused surfaces: union of constant mean curvature interfaces dictated by thin-film tension. \emph{Langmuir} \textbf{2021}, \emph{37}, 6601--6612\relax
\mciteBstWouldAddEndPuncttrue
\mciteSetBstMidEndSepPunct{\mcitedefaultmidpunct}
{\mcitedefaultendpunct}{\mcitedefaultseppunct}\relax
\EndOfBibitem
\bibitem[Grace(1982)]{grace1982dispersion}
Grace,~H.~P. Dispersion phenomena in high viscosity immiscible fluid systems and application of static mixers as dispersion devices in such systems. \emph{Chemical Engineering Communications} \textbf{1982}, \emph{14}, 225--277\relax
\mciteBstWouldAddEndPuncttrue
\mciteSetBstMidEndSepPunct{\mcitedefaultmidpunct}
{\mcitedefaultendpunct}{\mcitedefaultseppunct}\relax
\EndOfBibitem
\end{mcitethebibliography}

\end{document}